\documentclass[aps,prl,twocolumn,nopacs,superscriptaddress]{revtex4}

\usepackage{graphicx}
\usepackage{verbatim}
\usepackage{mathrsfs}
\usepackage{amsmath,amsfonts,amssymb}
\usepackage{graphicx}

\def\3{2.8in}    
\def\2{2.5in}
\def\4{3.0in}

\def \beq {\begin{equation}}
\def \eeq {\end{equation}}
\begin{document}

\title{Surface Versus Bulk Dirac States Tuning in a Three-Dimensional Topological Dirac Semimetal}


\author{Madhab Neupane}\affiliation {Joseph Henry Laboratory, Department of Physics, Princeton University, Princeton, New Jersey 08544, USA}

\author{Su-Yang Xu}\affiliation {Joseph Henry Laboratory, Department of Physics, Princeton University, Princeton, New Jersey 08544, USA}

\author{Nasser Alidoust}\affiliation {Joseph Henry Laboratory, Department of Physics, Princeton University, Princeton, New Jersey 08544, USA}

\author{Raman Sankar} \affiliation{Center for Condensed Matter Sciences, National Taiwan University, Taipei 10617, Taiwan}


\author{Ilya Belopolski}\affiliation {Joseph Henry Laboratory, Department of Physics, Princeton University, Princeton, New Jersey 08544, USA}

\author{Guang Bian}\affiliation {Joseph Henry Laboratory, Department of Physics, Princeton University, Princeton, New Jersey 08544, USA}

\author{Daniel S. Sanchez}\affiliation {Joseph Henry Laboratory, Department of Physics, Princeton University, Princeton, New Jersey 08544, USA}

\author{Chang Liu}\affiliation {Joseph Henry Laboratory, Department of Physics, Princeton University, Princeton, New Jersey 08544, USA}






\author{Tay-Rong Chang} \affiliation{Department of Physics, National Tsing Hua University, Hsinchu 30013, Taiwan}

\author{Horng-Tay Jeng} \affiliation{Department of Physics, National Tsing Hua University, Hsinchu 30013, Taiwan} \affiliation{Institute of Physics, Academia Sinica, Taipei 11529, Taiwan}

\author{BaoKai Wang}
\affiliation{Centre for Advanced 2D Materials and Graphene Research
Centre, National University of Singapore, 6 Science Drive 2, Singapore
117546}
\affiliation{Department of Physics, National University of Singapore,
2 Science Drive 3, Singapore 117542}
\affiliation{Department of Physics, Northeastern University, Boston,
Massachusetts 02115, USA}

\author{Guoqing Chang}
\affiliation{Centre for Advanced 2D Materials and Graphene Research
Centre, National University of Singapore, 6 Science Drive 2, Singapore
117546}
\affiliation{Department of Physics, National University of Singapore,
2 Science Drive 3, Singapore 117542}

\author{Hsin Lin}

\affiliation{Centre for Advanced 2D Materials and Graphene Research
Centre, National University of Singapore, 6 Science Drive 2, Singapore
117546}
\affiliation{Department of Physics, National University of Singapore,
2 Science Drive 3, Singapore 117542}

\author{Arun Bansil}\affiliation {Department of Physics, Northeastern University, Boston, Massachusetts 02115, USA}





\author{Fangcheng Chou} \affiliation{Center for Condensed Matter Sciences, National Taiwan University, Taipei 10617, Taiwan}



\author{M. Zahid Hasan}\affiliation {Joseph Henry Laboratory, Department of Physics, Princeton University, Princeton, New Jersey 08544, USA}
\affiliation {Princeton Center for Complex Materials, Princeton University, Princeton, New Jersey 08544, USA}


\begin{abstract}


{Recently, crystalline-symmetry-protected three-dimensional (3D) bulk Dirac semimetal phase has been experimentally identified in a stoichiometric high-mobility compound, Cd$_3$As$_2$. The Dirac state observed in Cd$_3$As$_2$ has been attributed to originate mostly from the bulk state while calculations show that the bulk and surface states overlap over the entire Dirac dispersion energy range. In this study, we unambiguously reveal doping induced evolution of the ground state of surface and bulk electron dynamics in a 3D Dirac semimetal. We develop a systematic technique to isolate the surface and bulk states in Cd$_3$As$_2$, by simultaneously utilizing angle-resolved photoemission spectroscopy (ARPES) and $in$-$situ$ surface deposition. Our experimental results provide a method for tuning the chemical potential as well as to observe surface states degenerate with bulk states, which will be useful for future applications of 3D Dirac  semimetal.}

\end{abstract}
\date{\today}
\maketitle

Two-dimension (2D) Dirac electron systems often exhibit exotic quantum phenomena and, therefore, have been considered as one of the central topics in modern condensed matter physics. The most promising examples are the extensive studies on graphene and the surface states of topological insulators (TI) \cite {Dirac, Graphene, RMP, Zhang_RMP, David_nature, Xia, David_science, Cava_PRB, Chen_Science, Suyang, Neupane_1}.
Dirac semimetals  have gapless linear dispersions in three spatial dimensions, which can be  regarded as a 3D analog of graphene. Recently, 3D Dirac semimetals have been theoretically predicted  \cite{Dirac_semi, Dai} and experimentally realized in Cd$_3$As$_2$ \cite{Neupane, Cava,Chen_Cd3As2, XJZhou} and Na$_3$Bi \cite{Chen_Na3Bi, Hasan_Na3Bi_1,Hasan_Na3Bi_2}. These Dirac semimetals possess two Dirac nodes which are protected
by the crystalline symmetry and are believed to hold the key for realizing exotic Weyl physics \cite{Chen_Cd3As2, Dirac_3D, Dirac_semi, 3D_Dirac, Dai, Weyl, Volovik, Fang, Ashvin, Balent, Neupane, Cava,Chen_Na3Bi, Hasan_Na3Bi_1,Hasan_Na3Bi_2,XJZhou}. It has been theoretically understood that a 3D Dirac node can be viewed as a composite of two sets of distinct Weyl fermions \cite{3D_Dirac, Dai}. Moreover, if time-reversal or space inversion symmetry is broken, then the theory predicts that the 3D Dirac point (DP) can split into two Weyl nodes that are separated in momentum space. Thus, a Weyl semimetal phase can be realized,  which can be described by Weyl nodes at the Fermi level for the bulk and the presence of the exotic spin- polarized Fermi arcs on the surfaces connecting the bulk nodes \cite{Dai}.

The Dirac state observed in Cd$_3$As$_2$ system has been attributed to originate, mostly, from the bulk states \cite{Dai, Neupane, Cava} where the Dirac dispersion varies with photon energy \cite{Neupane, Cava}. Also, spin-resolved angle-resolved photoemission spectroscopy (SR-ARPES) measurements show negligible spin polarization \cite{Neupane} suggesting its bulk origin. However, calculations show that the bulk and surface states overlap on the entire Dirac dispersion energy range \cite{Dai, Neupane}. Therefore, it is natural to ask  a question: how can the surface state be resolved in the Cd$_3$As$_2$ system? Furthermore, the overlapping of the surface and bulk bands may cause the hybridization between them, which makes it difficult to separate the contributions of the surface and bulk states. Hence, it is important to develop a technique to enhance the spectral weight of the surface states in Cd$_3$As$_2$. 


In this Letter, we report the realization of the doping induced evolution of the ground state of surface and bulk electron dynamics in a 3D Dirac semimetal. We develop a methodology to isolate the surface and bulk states in high-mobility stoichiometric 3D Dirac semimetal Cd$_3$As$_2$. Using high-resolution angle-resolved photoemission spectroscopy (ARPES) and \textit{in-situ} alkali metal surface deposition, we show that Cd$_3$As$_2$ features a bulk Dirac cone located at the center of the (001) surface Brillouin zone (BZ), whose DP is tunable across the binding energy. This surface doping induced tunability of the chemical potential reveals between the variation of surface versus bulk Dirac dispersion. 
The reported observation of a tunable bulk Dirac semimetal phase in stoichiometric Cd$_3$As$_2$ with a high Fermi velocity and electron mobility opens the door to observe exotic 3D Dirac relativistic physics in bulk materials and to realizing the exciting Weyl semimetal topological phase with novel surface Fermi arcs.

\begin{figure}
\centering
\includegraphics[width=8.5cm]{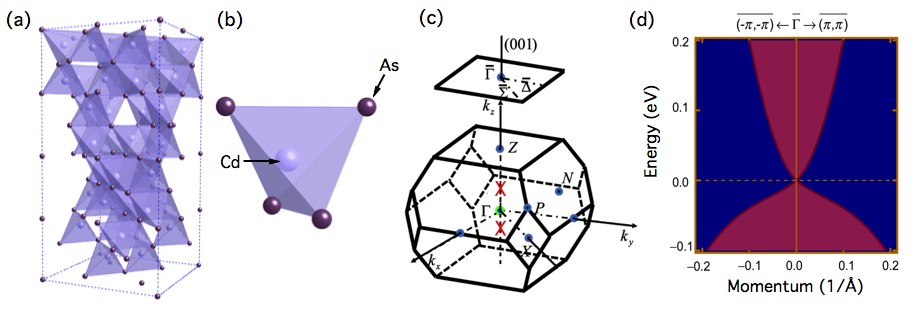}
\caption{{Crystal structure.} ({a}) Cd$_3$As$_2$ crystalizes in a tetragonal body center structure with space group of $I4_1$cd, which has 32 number of formula units in the unit cell. ({b}) The basic structure unit is a 4 corner-sharing CdAs$_3$-trigonal pyramid. {(c)} The bulk Brillouin zone (BZ) and the projected surface BZ along the (001) direction. The red crossings locate at $(k_x,k_y,k_z)=(0,0,0.15\frac{2\pi}{c*})$ ($c*=c/a$). (d) First-principles calculation of the bulk electronic structure. Projected bulk band structure on to the (001) surface, where the shaded area shows the projection
of the bulk bands.}
\end{figure}

Single crystalline samples of Cd$_3$As$_2$ used in this study were grown using the standard method, which is described elsewhere \cite{crys_str}. ARPES measurements for the low energy electronic structure were performed at the beamlines 4.0.3 and 12.0.1 at the Advanced Light Source (ALS) in Berkeley California, equipped with high efficiency VG-Scienta R8000 and R200 electron analyzers. Samples were cleaved \textit{in situ} and measured at $10$ K in a vacuum better than $1\times10^{-10}$ torr. The energy resolution was 10-30 meV and the angular resolution was better than 0.2$^{\circ}$ for all synchrotron-based ARPES measurements.
Samples were observed to be stable and, for a typical 20 hour measurement period, unsusceptible to degradation.
Sodium deposition was performed at the beamline 4.0.1 of the ALS from SAES getter source (SAES Getters USA, Inc.), which was thoroughly degassed before starting the experiment. During deposition, pressure in the experimental chamber was $<$ $1\times10^{-10}$ torr.  
First-principles calculations were based on the generalized gradient approximation (GGA) \cite{Perdew} using the projector augmented wave method \cite{Blochl} as implemented in the VASP package \cite{Kress}. The experimental crystal structure was used \cite{crys_str}. The electronic structure calculations were performed over $4\times4\times2$ Monkhorst-Pack $k-$mesh with the spin-orbit coupling included self-consistently.


\begin{figure*}
\centering
\includegraphics[width=18.5cm]{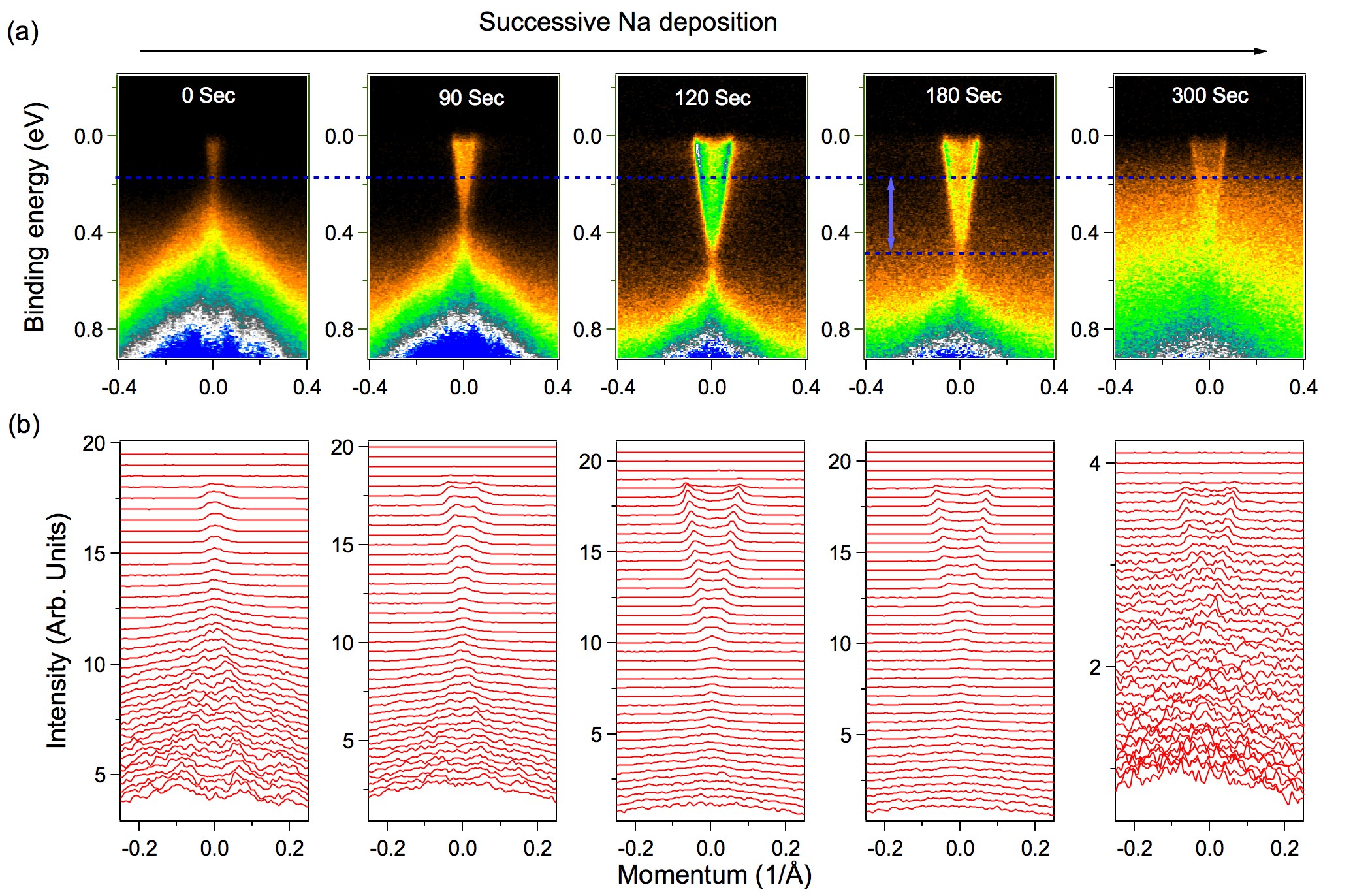}
\caption{{Fermi level tuning by surface Na deposition.} ({a}) ARPES dispersion map with successive Na deposition. The time for Na deposition is noted on spectra. Dirac-cone-like intensity continuum is observed for no surface deposition spectrum. Long (short) blue dash line shows the binding energy position of the Dirac point without (with) surface deposition. And vertical arrow indicates the energy shift of the binding energy ($\sim$ 300 meV). These data were measured with photon energy of 102 eV and temperature of 10 K at ALS BL4. 
(b) Corresponding momentum distribution curves (MDCs).}
\end{figure*}

We start our discussion by reviewing the crystal structure and the Brillouin zone (BZ) of Cd$_3$As$_2$. The crystal structure of Cd$_3$As$_2$ has a tetragonal unit cell with $a= 12.67$ \AA, $c= 25.48$ \AA, $Z= 32$, and the space group of $I4_1$cd (see Figs. 1a and b). In this structure, arsenic ions are approximately cubic close-packed and Cd ions are tetrahedrally coordinated. Each As ion is surrounded by Cd ions at six of the eight corners of a distorted cube (see Fig. 1b), with the two vacant sites located at the diagonally opposite corners of a cube face \cite{crys_str}. The corresponding BZ is shown in Fig. 1c, where the center of the BZ is the $\Gamma$ point, the centers of the top and bottom square surfaces are the $Z$ points, and other high symmetry points are also noted. In electrical transport measurements, Cd$_3$As$_2$ has long attracted much attention because of its very high electron mobility up to $\simeq10^5$ cm$^2$V$^{-1}$s$^{-1}$ \cite{Mobi} and interesting magneto-transport properties including a very small effective mass \cite{Magneto}. In theoretical calculations, Cd$_3$As$_2$ is of further interests because it is believed to have an inverted band structure \cite{Inverted}. More interestingly, a very recent theoretical study \cite{Dai} has shown that spin-orbit interaction in Cd$_3$As$_2$ cannot open up a full energy gap between the inverted bulk conduction and valence bands due to the protection of an additional crystallographic symmetry \cite{Dirac_3D} (in the case of Cd$_3$As$_2$ it is the $C_4$ rotational symmetry along the $k_z$ direction \cite{Dai}), which is in contrast to other band-inverted systems such as Bi$_2$Se$_3$ or HgTe \cite{RMP}. Theory predicts \cite{Dai} that the $C_4$ rotational symmetry protects two bulk (3D) Dirac band touchings at two special $\mathbf{k}$ points along the $\Gamma-Z$ momentum space cut-direction, as shown by the red crossings in Fig. 1c.  At the (001) surface, theoretical calculations show the presence of one 3D Dirac cone at the BZ center point ($\bar{\Gamma}$) as shown in Fig. 1d.
Therefore, Cd$_3$As$_2$ offers a platform for an exotic space group symmetry-protected bulk Dirac semimetal (BDS) phase, which additionally features a very high carrier mobility and an inverted band-structure.


In order to convey the tunability of the chemical potential, we systematically study the electronic structure of Cd$_3$As$_2$ at the cleaved (001) surface by successive Na deposition. 
High-resolution ARPES dispersion measurements are performed in the close vicinity of the Fermi level as shown in Fig. 2. The used photon energy of 102 eV corresponds to the photon energy at which bulk Dirac dispersion is observed. At this ``magic'' photon energy, a linearly dispersive Dirac cone is observed at the surface BZ center $\bar{\Gamma}$ point, whose Dirac node is found to be located at binding energy of $E_{\textrm{B}}\simeq0.2$ eV. At the Fermi level, only the upper Dirac band but no other electronic states are observed. On the other hand, the lower Dirac cone is found to coexist with a parabolic bulk valence band, which can be seen from Fig. 2a. Thus the data is consistent with recent results \cite{Neupane, Cava} and theoretical calculations  \cite{Dai}. In theory, there are two 3D Dirac nodes that are expected at two special $\mathbf{k}$ points along the $\Gamma-Z$ momentum space cut-direction, as shown by the red crossings in Fig. 1d. At the (001) surface, these two $\mathbf{k}$ points along the $\Gamma-Z$ axis project on to the $\bar{\Gamma}$ point of the (001) surface BZ (Fig. 1d). Therefore, at the (001) surface, theory predicts the existence of  one 3D Dirac cone at the BZ center $\bar{\Gamma}$ point, as shown in Fig. 2a. 

More substantial changes are observed by the \textit{in situ} surface deposition of alkali metal (sodium, Na) on the cleaved surfaces of  Cd$_3$As$_2$. As a function of Na-deposition time, three stages can be identified. At first for moderate Na deposition ($\sim$ 60 seconds) the Dirac point (DP) moves to higher binding energy by electron doping and a sharper state is observed to emerge at the edge of the bulk continuum. We can identify this stage at which surface states start to develop. For intermediate Na deposition [$\sim$ 60 to 120 seconds, Fig. 2 middle panel], the electron doping further increases and sharper boundary states appear. This stage corresponds with fully develop surface states.
Finally, heavy Na deposition [beyond
180 seconds] leaves the surface states and DP unchanged, which indicates that the sample cannot be further doped. These observations are further illustrated by the momentum distribution curves (MDCs) plotted in Fig. 2b.

In order to directly visualize the isolation of the surface and bulk states in Cd$_3$As$_2$, we compare MDCs extracted from 50 meV binding energy at different deposition time (see Fig. 3). It is worth noting that the topological Dirac SM carries a nontrivial 2D Z$_2$ invariant on either the k$_z$ = 0 or the k$_z$ = $\pi$ plane. Therefore, the Dirac semimetal turns into a strong Z$_2$ topological insulator if an external perturbation breaks the $C_4$ symmetry but does not break time reversal symmetry \cite{Nagaosa}. Thus the projected state on the (001) surface gives both bulk and surface states, as shown schematically in Fig. 3a. The MDCs at different deposition times are shown in Fig. 3c, whereas the white-dash line on the ARPES spectra in Fig. 3b depicts
the approximate energy position of the extracted MDCs.
From Fig. 3c, it is evident that after deposition the double peaks emerge from a single peak of MDC before deposition.  These two peaks correspond to the surface state of linear dispersion. 

In order to illustrate the surface state nature of the dispersion after deposition, we perform photon energy dependent ARPES measurement as shown in Fig. 4.  For this purpose, we use a relatively $p$-type sample where the chemical potential is located in the vicinity of the DP (see Fig. 4). A 3D Dirac semimetal is shown to feature nearly linear dispersions along all three momentum space
directions close to the crossing point, even though Fermi velocity can vary significantly along different directions \cite{Neupane}.
With the introduction of $in$-$situ$ Na deposition of $\sim$ 120 seconds, the chemical potential is shifted up by 300 meV, leaving the upper cone is now visible. Furthermore, the boundaries of the upper dispersion maps seem to be sharper. To identify their origin, we perform photon energy dependent measurements from 96 to 110 eV with a 2 eV energy increment (see Fig. 4). Upon varying the photon energy, the resulting dispersion maps are found to be unchanged, which is evidence supporting their 2D nature (see Fig. 4). 
These measurements further support our claim, which is that the precise control of the carrier density, and, therefore, the desired Fermi level can be
achieved in the Cd$_3$As$_2$ system


Now we discuss some interesting consequences of our experiment. First, the surface alkali metal deposition on Cd$_3$As$_2$ has revealed that the chemical potential of the bulk Dirac point can be tuned, at least near the surface since ARPES is relatively surface sensitive. The resulting measurements of this study show that the Dirac point is shifted in binding energy by more than $\sim$ 300 meV. The ability to tune the chemical potential make this compound interesting for transport measurements and, just as importantly, a promising candidate for future experimental studies on exotic phases.
Second, before the surface deposition, the bulk Dirac cone shows intensity filled inside the cone. The outer-boundary of the bulk Dirac cone dispersion, which mostly represents the surface states, is observed to become brighter and sharper after Na deposition (see Fig. 2 and Fig. 4). Thus, the deposition of the alkali metal on the cleaved surface of Cd$_3$As$_2$ can serve as a methodology to isolate the surface and bulk bands in this system.
Third, our photon energy dependent measurements reveal that the surface deposition provide a technique for tuning the bulk versus surface state. By the addition of Na on the surface of Cd$_3$As$_2$, the ARPES signal becomes more surface like which may probably be due to decreasing the penetration depth of the photon flux.

\begin{figure}
\centering
\includegraphics[width=8.9cm]{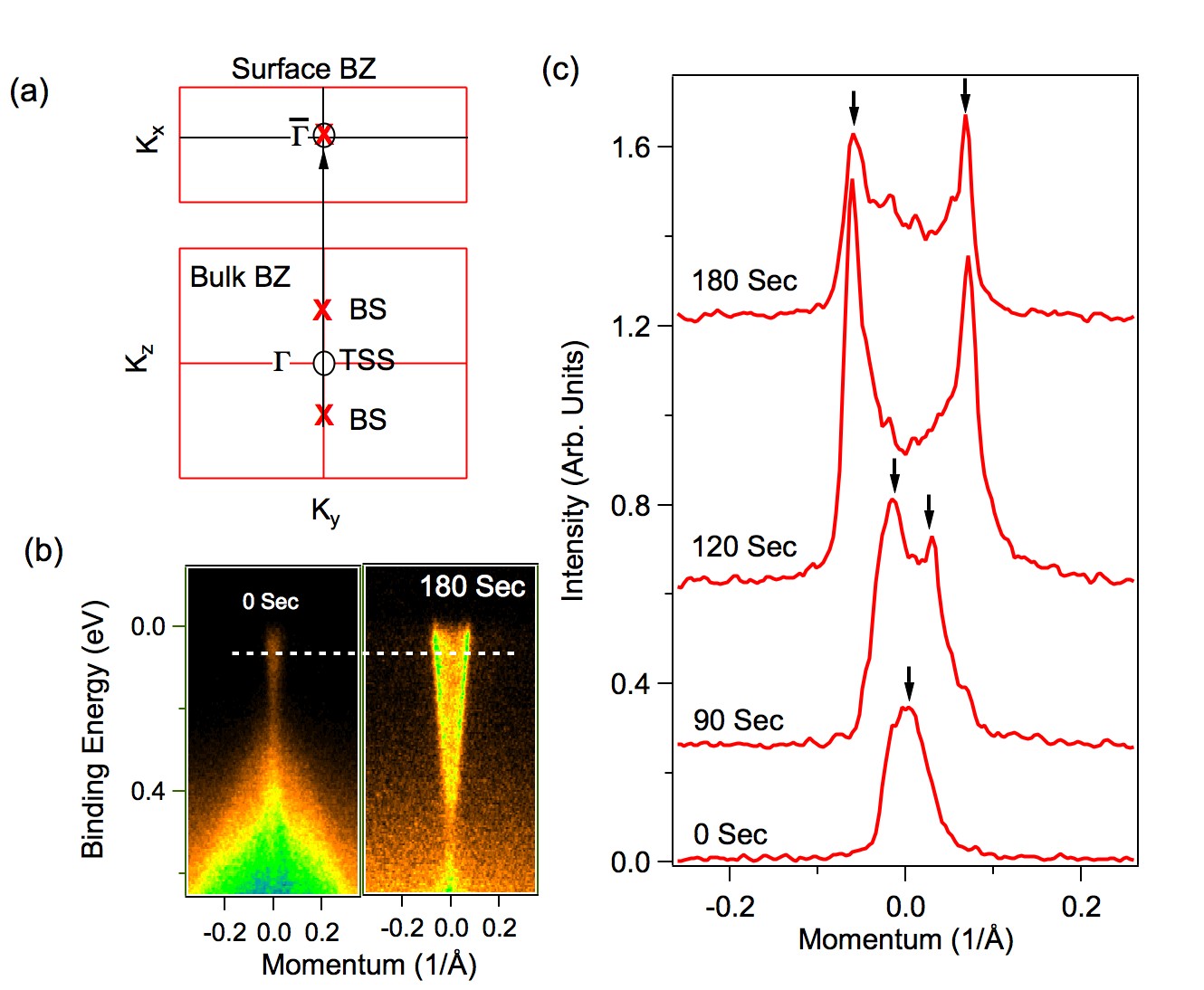}
\caption{{Surface vs bulk state tuning.} ({a}) Schematic view of the surface and bulk BZ along (001) plane. The red x sign shows the bulk Dirac cone and circle sign denote the topological surface state. BS and TSS represent bulk Dirac state and topological surface states. ({b}) ARPES measured surface dispersion maps of Cd$_3$As$_2$. The deposition time is noted on the spectra. The white dash-line represents the approximate binding energy position for the extracted momentum distribution curves (MDCs). ({c}) MDCs extracted from the spectra measured with different deposition timesz. The arrow lines represent the peak position. The evolution of the double peaks from a single peak shows how the spectral weight mostly dominated by surface states arise from the bulk state.}
\end{figure}

\begin{figure}
\centering
\includegraphics[width=8.9cm]{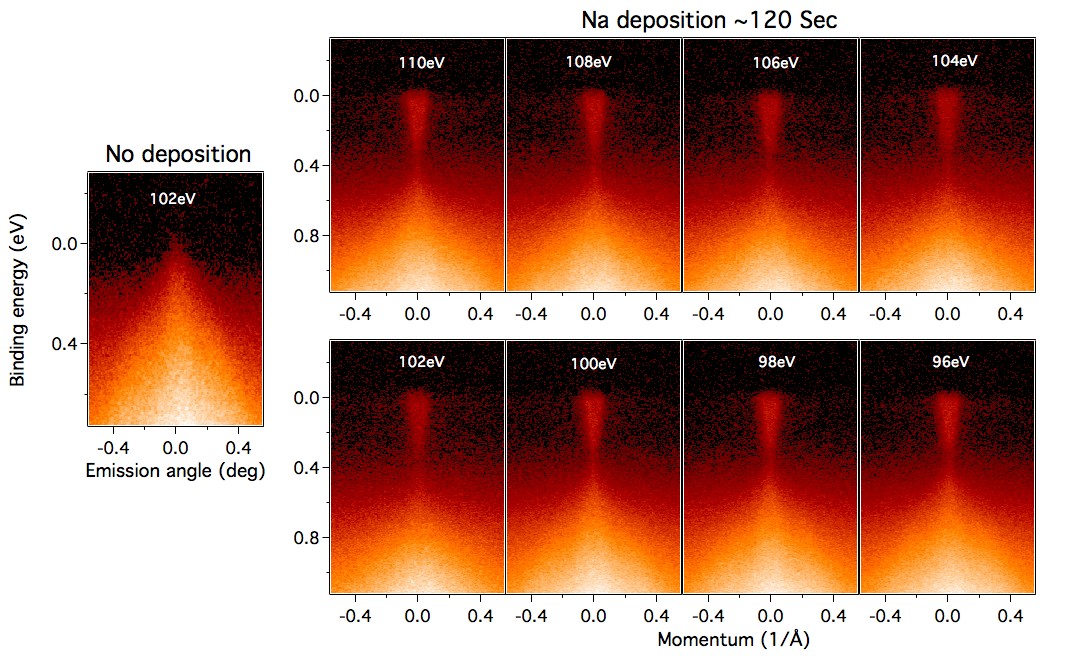}
\caption{{Observation of surface state.} 
ARPES dispersion maps measured before and after Na deposition. 
The sample measured here is relatively $p-$type, which is prepared under slightly different growth conditions (temperature and growth time).
Photon energy dependent ARPES dispersion maps after 120 seconds of Na deposition. The measured photon energies are noted on the spectra.}
\end{figure}


In conclusion, we experimentally show how to tune the chemical potential of the three-dimensional Dirac semimetal compound Cd$_3$As$_2$ by simultaneously executing surface alkali metal deposition and ARPES measurements.
In doing so, we develop a methodology to distinguish between the surface and bulk states in this high-mobility compound. The observation of the tunable 3D bulk Dirac phase and the surface doping induced surface versus bulk tunability in Cd$_3$As$_2$ paves the way for realizing a number of exotic topological phenomena and raises its potential application for future devices.

\bigskip
\textbf{Acknowledgements}
\newline
Work at Princeton University is supported by the US 
National Science Foundation Grant, NSF-DMR-1006492. 
M. Z. H. acknowledges visiting-scientist support from 
Lawrence Berkeley National Laboratory and additional 
partial support from the A. P. Sloan Foundation and 
NSF-DMR-0819860. 
The work at Northeastern University was supported by the US Department of Energy (DOE), Office of Science, Basic Energy Sciences grant number DE-FG02-07ER46352, and benefited from Northeastern University's Advanced Scientific Computation Center (ASCC) and the NERSC supercomputing center through DOE grant number DE-AC02-05CH11231.
H. L. acknowledges the Singapore 
National Research Foundation for support under NRF 
Grant No. NRF-NRFF2013-03. T.-R. C. and H.-T. J. are 
supported by the National Science Council, Taiwan. 
H.-T. J. also thanks NCHC, CINC-NTU, and NCTS, 
Taiwan, for technical support. 
We thank J. D. Denlinger and A. V. Fedorov for beamline assistance at
the Advanced Light Source (ALS-LBNL) in Berkeley.
M.N. acknowledges discussion with  A. Alexandradinata
and Z. Wang from Princeton Physics.

\newpage

\noindent

%
%
%

\end{document}